# Tokenising behaviour change: optimising blockchain technology for sustainable transport interventions

Iain Barclay[1*], Michael Cooper[2+], Alun Preece[1],
Omer Rana[1] and Ian Taylor[1,3]

[1] School of Computer Science and Informatics, Cardiff University, Cardiff, UK
[2] Emergence LLC, Denver, CO, US
[3] Center for Research Computing, University of Notre Dame, IN, US
*BarclayIS@cardiff.ac.uk +emergence.cooper@gmail.com

**Abstract.** Transport makes an impact across SDGs, encompassing climate change, health, inequality and sustainability. It is also an area in which individuals are able to make decisions which have potential to collectively contribute to significant and wide-ranging benefits. Governments and authorities need citizens to make changes towards adopting sustainable transport behaviours and behaviour change interventions are being used as tools to foster changes in travel choices, towards more sustainable modes. Blockchain technology has the potential to bring new levels of scale to transport behaviour change interventions, but a rigorous approach to token design is required. This paper uses a survey of research projects and use cases to analyse current applications of blockchain technology in transport behaviour change interventions, and identifies barriers and limitations to achieving targeted change at scale. The paper draws upon these findings to outline a research agenda that brings a focus on correlating specific Behaviour Change Techniques (BCTs) to token design, and defines processes for standardising token designs in behaviour change tools. The paper further outlines architecture and operational considerations for blockchain-based platforms in behaviour change interventions, such that design choices do not compromise opportunities or wider environmental goals.

## 1 Introduction

The United Nations Sustainable Development Goals (SDGs) define a set of objectives that together represent a global call to action to bring an end to poverty, protect the planet and ensure that all people are able to enjoy peace and prosperity by 2030 (UN General Assembly, 2015). A need to focus on behaviour change to achieve the SDG's has been documented by prominent actors in this space, including the World Bank (2014). Behaviour change at this scale is a daunting task, but the emergence of blockchain technology and the opportunities it creates for the use of digital tokens as behaviour change tools, could play a critical role towards meeting the SDG's. Such a scenario requires building on current use cases of blockchain technology to achieve targeted behaviours, and avoid undesired behaviours, using digital tokens to help align incentives around achieving specific outcomes. Populations across the globe experience the digital revolution in different ways, but these early use cases can be combined with current research in identifying specific token designs that are optimal for behaviour targets, to develop appropriate and scalable behaviour change programmes.

Transport plays a part in many of the SDGs, including the promotion of good health and well-being for all (SDG 3), combating climate change and its implications (SDG 13), reducing inequality (SDG 10) and making cities inclusive, safe and sustainable (SDG 11) (Macmillan, et. al, 2020). Token-based interventions are a well-established approach in travel behaviour change (Everett, Hayward and Meyers, 1974), and blockchain technology provides a solid technical platform for implementation of token economies (Lee, 2019). Exploring the future role that tokenised blockchain-based systems can play in fostering momentum towards increased adoption of sustainable transport across the globe provides a helpful domain for research, extending from behaviour science through operationalisation of blockchain-based deployments.

In this paper we use studies from the transport arena to motivate an analysis of blockchain adoption in token-based behaviour change interventions. Section 2 provides an initial overview of tokens, their history as behaviour change tools, and their potential when coupled with blockchain technology. The role that transport plays across multiple SDGs is expanded upon in Section 3, which discusses the important contribution that behaviour change techniques (BCT) can make in increasing adoption of sustainable transport modes in order to meet the objectives of the SDGs. Section 4 presents a survey of early research projects which are using tokens and blockchain technology to target specific travel behaviours. The current state of token research is summarised in Section 5, which presents the main contribution of this paper in outlining a research agenda for taking blockchain tokens to scale as behaviour change tools for achieving relevant SDGs. This agenda identifies the need to integrate

BCTs into token taxonomies, and develop definition of a level of standardisation for token design to allow for evaluation and comparison and operational frameworks for token design. Section 6 introduces a conceptual architecture for blockchain-based behaviour change systems, and discusses considerations for operationalising token-based schemes.

## 2 Tokens as Behaviour Change Tools

Token economies, which provide users with a reward for adopting targeted behaviours, are part of the learning approach of Operant Conditioning, pioneered by B F Skinner (Catania and Harnad, 1988). Typically a token economy intervention rewards the subject with an allocation of a token (secondary reinforcer) for each instance of the targeted behaviour they undertake. The token economy provides a means to enable the subject to later exchange accumulated tokens for a reward (primary reinforcer), typically something the subject desires, chosen from a selection of alternatives. Token economies have been used as behaviour change interventions over several decades (Ayllon and Azrin, 1968) and have been found to be successful in delivering sustained changes in behaviour, even after the removal of the incentive program (Doll, McLaughlin and Barretto, 2013).

Historical usage of tokenised incentives is limited in application, primarily focussing on education and health, and in scope. However, the digitization of societies has contributed to new opportunities to expand use of tokenised schemes and develop the resulting evidence base. Distributed digital ledger and blockchain technologies, for example, have introduced new opportunities to use tokens in novel and cost-effective ways, with greater ability to experiment and adapt. Blockchains, as implementations of distributed ledgers, provide tamper-proof recording of digital values, or balances, in a decentralised and transparent environment. Blockchain platforms such as Ethereum which offer programmable smart contracts (Antonopoulos and Wood, 2018) can perform autonomous transactions on assets represented on a blockchain, supporting the transfer of tokenised assets from one party to another, and the conversion of tokenised assets from one form to another. As such, blockchain technologies have the potential to provide a robust and secure mechanism for the implementation and delivery of token-based systems and token economies, and are well suited to environments where there is no inherent trust between the participants (Lee, 2019).

More precisely, blockchain technologies provide an immutable record of transaction histories which can be used as part of a technology platform to keep a record of interactions between parties. Blockchain consensus mechanisms ensure that this record is accurate and cannot be subsequently be revised or changed by any party. This property means that blockchain technology can be used to build a trusted record of events into a system, even when the participants in the system are not of equal standing. In a token economy, this would provide a means to maintain an accurate, and unchangeable, record of tokens issued and exchanged between participants.

As a general use technology, blockchains have been used to tokenize a wide variety of ecosystems to solve a variety of behavioural problems. The scope of these applications has led to different classes and designs of tokens correlated to specific types of functions, including those encompassing transportation issues.

## 3    Targeting Transportation Behaviour Change

Macmillan, et al, (2020) provide compelling, evidence-based links between impacts resulting from reduced use of private car travel and multiple SDG outcomes, a shift which can come about by fostering an increase in walking, cycling and other personal travel changes towards active or sustainable travel. Presenting a systematic literature review on transport behaviour change interventions, Roy, et al (2021) determine that shifting travel towards active modes can help to generate income for the local communities (SDG 1 and 8) and provide commuters with low-cost methods for accessing basic services (Macmillan et al., 2020) particularly in developing countries, and can reduce gender inequities in gaining access to basic services, including healthcare and education (SDG 5).

At a more local level, many regions, cities and towns have developed their own climate and mobility targets (Clarke and Ordonez-Ponce, 2017), often based on the SDGs. These local targets can be challenging to meet, Foster, Lamura and Hackel (2020) describe the Austrian capital city of Vienna, for example, as being unable to meet their targets without the active involvement of the citizens of the city. Without engaged citizens many targets for climate change will be missed, and without citizens themselves getting active, health and mobility targets will also go by the wayside, which could lead to significant – and potentially costly -- physical and mental health and well-being implications.

An ongoing approach that city authorities and other stakeholders can adopt as they tackle climate and mobility targets is to develop and deploy interventions that seek to change the behaviours of their citizen. Through appropriate design of behaviour change interventions, agencies can move to direct citizens towards using sustainable travel modes, such as taking public transport where available, or adopting active travel modes such as walking and cycling (Scheepers et al, 2014).

Using incentives, in the form of direct financial rewards or as rewards in a token economy, is proposed as a behaviour change intervention to encourage travellers to forsake car trips and adopt more sustainable modes of travel for some journeys. Recently, for example, Ricci et al. (2020) surveyed a group of 686 adults in Italy and found that 75.1% expressed that they would make changes to their journey patterns in order to receive an incentive, with discounts on their energy bills reported as the most popular reward.

## 4     Leveraging Behaviour Change in Travel Mode Choices

A literature review identified four projects which have proposed or are making use of blockchain technology to try to increase adoption of sustainable transport by individuals, and in particular to encourage travel by means of cycling and walking.

Of the projects analysed (Table 1), one is currently deployed in a pilot project, one is at a smaller scale field trial stage. The other projects are no longer thought to be active. Three of the analysed research projects are implementations of token economies, which use blockchains to record and maintain token balances or to facilitate exchange of earned tokens as rewards. One of the projects further proposes to offer rewards for distance cycled and also provide a market for data gathered during the journey. The final project proposes a different mode of blockchain use, presenting early-stage concepts intended to make use of blockchain smart contracts to encode commitments related to making walking journeys. The use of smart contracts to encode commitments could be integrated into a token economy and used to autonomously issue earned rewards.

| Project | Status | Model |
|---|---|---|
| Kultur-Token | Pilot | Rewards earned in token economy |
| Cycle4Value | Field Trial | Rewards earned in token economy |
| Geolocation | Inactive | Token economy and data market |
| Walkers Union | Inactive | Social commitments via smart contract |

Table 1 – Summary of projects in survey

The analysed research projects all make use of behaviour change tools relevant to current token research, including:

- Reinforcing positive new beliefs.
- Shaping emerging habits with new offerings.
- Sustaining new habits, using contextual cues.
- Aligning messages to consumer mindsets.
- Analysing consumer beliefs and behaviours at a granular level.

Whilst the detail of implementation of each approach varies, the schemes which are most progressed typically make use of a customised smartphone app which attempts to quantify the user's travel behaviour. This quantification is achieved by leveraging the smartphone's sensor technology to learn about the user's location, speed, acceleration, etc. and invoking AI services to make educated guesses about the user's travel mode and distance travelled. These estimated behavioural characteristics are then converted into tokens that represent the traveller's estimated carbon saving, compared to making the same journey by car. Tokens are accumulated as the user continues to use the app, and at some stage the user is able to convert their tokens for a reward of a different kind, depending on the individual service implementation. Figure 1 shows a schematic of the approach, showing interaction between the "Travel Detection" app element, issuance of tokens based on estimated travel, with token storage either in the app itself or held on the cloud in an account setup for the user.

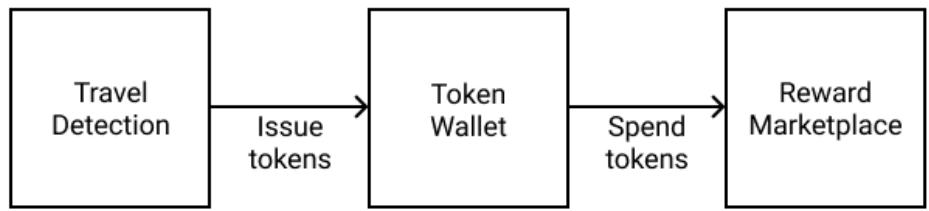

Figure 1. Schematic of mobility converting token economy

The final component is a marketplace, which shows the user items or services that can be purchased with their tokens. The service will provide mechanisms to exchange earned tokens for the chosen reward and adjust the user's token balance accordingly.

**4.1 Kultur-Token**

Kultur-Token (KT)[1] is a travel behaviour change intervention from Vienna, Austria. Travellers use a smartphone app which offers rewards for engaging in low-carbon travel behaviour. The KT app uses sensors in the traveller's smartphone to track their mobility behaviour, and rewards sustainable travel with tokens giving free access to cultural attractions in the city (Foster, Lamura and Hackel, 2020). KT has been described as "the first app to tokenize low-carbon mobility linked to citizen engagement and a vibrant cultural life"[2].

Foster, Lamura and Hackel (2020) describe the KT app detecting different travel modes - car, bicycle/scooter, walking and public transport. The app estimates the amount of $CO_2$ saved by adopting a low-carbon alternative to the car. Travellers are rewarded with a portion of a Kultur-Token for $CO_2$ savings made on non-car journeys. As low-carbon journeys are made the KT portions add up to a full token, representing 20kg $CO_2$ saving compared to using an average car. The traveller can exchange one full KT for a ticket to a participating cultural venue, which include the Concert Hall, a visual arts venue, and Vienna museum's 19 locations (Foster, Lamura and Hackel, 2020).

---

[1] https://digitales.wien.gv.at/site/projekt/kultur-token/ (In German)
[2] https://www.clicproject.eu/kultur-token-sustainable-business-model-visualizing-tokenizing-and-rewarding-mobility-behavior-in-vienna-austria/

Foster, Lamura and Hackel (2020) describe the KT app's three main screens. The first categorises the traveller's journeys as percentages travelled by walking, bicycle or scooter, and public transport. The screen shows the annual $CO_2$ saving as well as the aggregated saving for all participants. The second screen shows a pie chart, depicted as petals of a flower, as a visualisation of $CO_2$ saved by the app user. The petals fill as the user travels in a low carbon way, representing their $CO_2$ savings. A completed flower represents 20kg of saved $CO_2$, worth 1 KT. The third screen is the Marketplace, where the traveller can exchange a KT for an entrance ticket to one of the cultural institutions. Travellers select the venue they want to visit, and their token is spent, with a QR code presented in return. The QR code is scanned at the venue entrance for admission. Additionally, a tree is planted for every converted token (Foster, Lamura and Hackel, 2020). Conversion of KT tokens into the QR code ticket that provides entry into a venue is mediated by a blockchain platform (in the current pilot deployment generated tokens are not stored on a blockchain implementation). Mindful of the environmental impact associated with some blockchain mechanisms, and notably those that like Ethereum which use a proof-of-work consensus mechanism (Fairley, 2018), the KT project team chose to adopt a permissioned blockchain, which is based on a custom adaption of Ethereum and uses an alternative proof-of-authority for consensus.

In use, the app reads data from the smartphone's sensors and uses mobility-related information including GPS coordinates, speed, and acceleration to identify the journey type. $CO_2$ savings are estimated based on making the same journey by car. The app is reported to correctly detect travel mode and distance travelled with "an average accuracy of 90%" (Foster, Lamura and Hackel, 2020).

In January 2020, Austria's submission to the UN's Commission on Science and Technology for Development[3] stated the goal of the KT project as to "incentivize citizens of the City of Vienna, Austria, towards sustainable mobility behaviors that reduce greenhouse gas emissions associated with cars.". The reward itself, an entrance ticket to a cultural attraction, has been chosen specifically to meet other objectives from the City's agenda - to increase social inclusion and access to culture (Foster,

---

[3] https://unctad.org/system/files/non-official-document/CSTD_2020-21_c01_HB_Austria_en.pdf

Lamura and Hackel, 2020). The intervention forms part of Vienna's approach to meet a trio of objectives from the "Smart City Wien" program[4]. These objectives have targets for carbon reduction, increased mobility and more inclusive access to culture.

Using a digital infrastructure to estimate each user's CO2 savings and track this in the app, provides a mechanism to place a value on carbon savings. A process of tokenisation (Hargrave, Sahdev and Feldmeier, 2019) facilitated by the technology platform exchanges saved CO2 for rewards - effectively, the user's mobility pays for their cultural experience. As a result, City authorities can leverage citizen's activity, counting their contribution towards three objectives - mobility, carbon reduction and cultural activity. Foster, Lamura and Hackel (2020) describe the City's inability to meet targets without participation and engagement from citizens and identify KT as a digital technology providing an interface between the citizens and the policy goals. The depiction of "the city as an operating system" (Tomitsch, 2016, p.87) is appropriate - with Vienna and its infrastructure providing the platform upon which citizens act.

Foster, Lamura and Hackel (2020) suggest that KT belongs to a class of "sustainability behavioural change apps" (SBCA), adopting techniques that take advantage of smartphone sensors and near-ubiquitous internet availability to track and analyse activities, and present personalised feedback and information. SBCA apps are designed to modify behaviour through targeted information provision, incentives and gamification, which is discussed in a travel behaviour context by Yen, Mulley and Burke (2019). The KT app is based on the Changers[5] app, praised by Sullivan et. al, (2016, p.13) for providing real-world representation of CO2 savings, e.g. "this is equivalent to one cycle of laundry at 60°C" in a way that can be visualised by users.

1000[6] participants were selected at random to take part in the KT pilot in Vienna. Unfortunately, the pilot was suspended in March 2020 with the onset of the Covid-19 pandemic but is scheduled to resume in mid-2021.

---

[4] https://smartcity.wien.gv.at/en/

[5] https://changers.com

[6] https://www4.baumann.at/kultur-token-wien-testbetrieb-gestartet/ (In German)

In spite of the postponement, there are interesting aspects to the intervention that make it worthy of attention. Most notably, the mechanics of the intervention are cleverly designed, linking mobility and climate objectives to social objectives, powered by the choices made by travellers. As such, social co-benefits of improved cultural access are inherently plugged in to the intervention. It would be interesting to extend the intervention to allow travellers to donate earned tickets to charities, which might serve to further increase social inclusion and access to culture. It would be valuable to learn whether travellers are motivated to change travel behaviour by the promise of tree planting, the cultural attractions or simply enjoyment of participating in quantification and visualisation of their travel behaviour.

The intervention's design is elegant, although the incentive of a ticket to a cultural attraction does not seem likely to appeal to all, especially with an estimated two weeks of low-carbon travel required to earn it (Foster, Lamura and Hackel, 2020). There is a possibility that people most likely to use a smartphone-based mobility tracker are those who already use sustainable transport or make good use of cultural attractions, so social inclusion may not be increased as much as hoped. It would be interesting to see donation of rewards to charities or other groups offered in a future version of the pilot study, in order to attempt to increase social inclusion and access. Further, the additional reward of tree planting feels somewhat ancillary, and it appears there is no indication in the app to show how many trees have been planted. The app tracks mobility, so citizens are rewarded for all low carbon travel, not just journeys where car use has been avoided. This may be an inevitable lack of precision resulting from automated activity tracking but has the consequence of potentially over-estimating carbon savings contributed by the app – as some tracked journeys may always have been made on foot or on bicycle, and not form part of a travel mode change.

As a city-scale travel change intervention, the KT system's design has much to admire, and its coupling of travel mode choices to rewards that have the capability to contribute towards additional targets and objectives of the City provides a compelling model for other cities to study.

### 4.2 Cycle4Value

Cycle4Value[7] (C4V) seeks to increase bicycle use by providing incentives for cycling, manifested through the issuance of "Cycle Tokens" which can ultimately be traded in a marketplace for discount codes and other rewards. Seewald, et al (2021) describe the project as it is designed for a proof-of-concept, which is planned to lead to a field trial in cities in Austria and Germany.

The project team and supporting stakeholders have calculated cost savings (or associated financial benefits) of cycling grouped across themes of ecology, economy and health. Within these groups, the cost savings are further broken down. In Ecology, a price is placed on the benefits associated with a reduction in emissions of pollutants, particularly $CO_2$ and NOX; in Economy, on savings in individual costs as a result of cycling rather than using a car, and in Health on improvements in physical and mental health, and from improved traffic safety. The fiscal benefits of cycling are calculated in comparison with the same journey being made by car, as seen in other projects described in this review paper.

C4V uses an energy efficient blockchain platform[8] to generate tokens representing cycling journeys and issue them to cyclists. Seewald, et al (2021) describe the remuneration scheme as being the square root of cycling kilometres per tracked journey (with a limit of 4 tokens per journey) and an overall limit of 8 tokens issued to any one user in a day. Tokens are described as "semi-monetary compensation" as they have the potential to be converted to cash (by reselling rewards exchanged for tokens earned). As such, the C4V team are researching approaches to ensuring that the system is robust to attack and enumerate possible attack vectors from errant users attempting to increase their token balance: uploading tracks from sports events, uploading tracks from delivery services, uploading tracks recorded in other vehicles, using multiple phones on a bicycle, uploading generated fake tracks and uploading modified tracks. The team are seeking to mitigate against these attacks in a number of ways, and plan to host a competition to encourage friendly attacks to be made on the system in order to identify further issues. The remainder of the discussion of the C4V project presented by Seewald, et al (2021) discusses details of experiments in providing robust detection of genuine journey tracks, using a number of AI techniques. Treating the Cycle Tokens as having a fiscal value and providing strong measures to ensure that they cannot be earned erroneously is a unique and important part of the C4V project and will be of great benefit to the wider community when adopted more broadly.

---

[7] https://www.cycle4value.at/en/

[8] https://www.jelurida.com/ardor

**4.3   Monetizing Geolocation Data**

Jaffe, Mata and Kamvar (2017) propose a technical design based on the Ethereum blockchain platform. The proposed solution seeks to reward urban cycling through provision of financial incentives, funded by sponsorship from private companies, such as health insurance companies, governments and other organizations who in turn stand to gain benefit from an increase in cycling through better health of the population, or reduced traffic congestion and pollution. Ultimately, the paper's authors envisage the creation of a data exchange platform, where cyclists will be equipped with sensors to autonomously gather geospatial and environmental data, such as air quality and traffic congestion levels, as they travel. The platform will facilitate a marketplace for this data, leading to "a scalable micro-entrepreneurship ecosystem that promotes the use of sustainable urban transportation", the adoption of blockchain technology to build the platform seeks to provide a robust and transparent system, without requiring involvement of a trusted intermediary.

A proof-of-concept for the system enabled deployment of bicycle mounted sensors, and an Ethereum light client running on a Raspberry Pi device with additional GPS/GSM module to support user location and activity tracking and connectivity to the blockchain network, with a smart contract representing each user of the system and holding their earned funds.

This early project presented interesting ideas that continue to hold merit some years on. Deployment of a practical variant of this solution would need to remove the need for custom hardware, unless the user was provided this as part of their onboarding into the ecosystem. As such, the technological autonomy of the platform provided by use of a blockchain platform would need to be matched by an organisational infrastructure and governance infrastructure to provide hardware and operational support. The discourse on blockchain-governed organisations -- so called Distributed Autonomous Organisations, or DAOs -- has developed significantly (El Faqir, Arroyo and Hassan, 2020) since publication of Jaffe, Mata and Kamvar's paper (2017), and it would be very interesting to consider the role of a DAO in provisioning and running an active travel incentive program on a city scale. A community owned and operated organisation managing a viable sustainable travel scheme would seem to be a good fit.

### 4.4 Walker's Union: Making Contractual Commitments

As an alternative to using blockchain to provide tokenised incentives for activity Jiang, Giaccardi and Albayrak (2018) propose an approach to motivate walking by taking advantage of the smart contract capabilities of some blockchain platforms. Their proposal suggests four different modes of 'walking contract'. The first two contacts are considered social contracts, and consider people walking with a companion on the same route, or a remote companion, who shares the time and walking distance, but walks in a different location. The other proposals are described as environmental contracts and provide a means for the user to make commitments to non-human entities, which include a physical location, a pet, or digital content.

Considering each in turn, the social contract is suggested as a facilitator, to provide a means for walkers to express their availability, intended route and topics for conversation. The system intends to offer a matchmaking service and proposes to invite participants to walk together based on their expressed preferences. More interestingly, the second variant allows the system to connect walkers from different locations and enables them to walk "together", albeit remotely, with an online chat mode suggested. The environmental contract proposals are even less well developed but suggest that the user could make commitments to their environment, or to pets, for example, to walk to particular locations. A notion of an incentive for fulfilling a contract is introduced, as shopkeepers could reward walkers for walking to their shop, or media producers could offer rewards for consuming particular content whilst walking.

It is not apparent that this work has been further developed, but the use of smart contracts to provide digitally signed evidence of a commitment to partake in an activity presents interesting future possibilities, especially when paired with a social aspect of making a commitment to a third-party for a shared walk, either in-person or remotely. Participants in the economy could, for example, challenge each other individually or in groups, and smart contracts could autonomously enact the agreed actions at the conclusion of the challenge – a transfer of tokens from participants to the challenge winner, for example.

## 5 A Research Agenda for Tokenised Behaviour Change

There is an emerging field of research into the use of tokens that seeks to correlate the design of tokens to the problems they are intended to solve. A primary argument of this paper is that this research does not take behaviour science into sufficient account, meaning that the current research does not tightly correlate token design to behaviour change techniques. To better understand this need, an overview of the current fields of research is necessary.

### 5.1 Current Status of Token Design Research

A token can be simple or complex, much like the outcomes it seeks to leverage. The tokens outlined in the above research projects are relatively simple in that they are meant to carry out well-defined transactions that are dependent on readily verifiable conditions, such as distance and mode of travel. The tokens mentioned above are not managing complex ecosystems where a variety of actors are required to perform complex behaviours in shifting conditions. Taking tokenised ecosystems to the next level will require further development of a systematic approach to token design which tightly integrates behaviour change targets and barriers.

The field of blockchain token design is becoming known as Token Engineering[9] and is comprised of a collaboration of expertise and skill sets encompassing computer science, data science and social science (Voshmgir and Zargham, 2020). Design protocols are being established to break token design down into its various components (objectives, risks, value propositions, etc.) that can be envisioned as dials on a control board that can be modified if the results are not optimal (Dhaliwal, et al, 2019). Due to the various components of token design and the low barriers to modifying that design (by changing the code) based on measured results, there is huge increase in the ability to experiment. The primary barrier to optimal experimentation being the ability to correlate the contribution of the token design to the end behavioural outcome in a way that provides the evidence to inform the changes to the token design.

---

[9] https://tokenengineeringcommunity.github.io/website/

The desire to measure the effects of tokens on the environments in which they seek certain behaviours (i.e. measuring the token's contribution) has led to development of the field of token economics, known as "Tokenomics", and the emergence of an original evidence base on "what works". Additionally, token taxonomies are being developed that establish a spectrum of different token design types based on their function - whether it be to tokenize an asset or to provide decentralized finance products and services, for example[10].

## 5.2   Future Research Agendas

Behavioural sciences require measures that unpack the mechanisms which leverage behaviour. This requires designing the intervention in a manner that facilitates such measures. As digital tokens grow in scope and complexity, measuring these mechanisms (as opposed to solely measuring the outcomes they target) will become more critical. The history of tokens as behaviour change tools is a history steeped in evidence built on measuring these mechanisms. As tokens are increasingly digital and managed via blockchain platforms there is a need to build a new evidence base which necessitates measures of mechanisms and adapting the token design accordingly.

A focus for taking tokens as behaviour change tools to a larger scale is to implement a research agenda that accomplishes the following objectives:

- Integrating targeted behaviour changes into token taxonomies

---

[10] https://github.com/InterWorkAlliance/TokenTaxonomyFramework

- Defining a level of standardization that allows for use cases to yield a rigorous evidence base on "what works"
- Developing operational frameworks that outline the processes and protocols for different skill sets and functions for a token design and adaptation lifecycle

These points are expanded in the discussion that follows. A further area of research interest, which is discussed in Section 6, is in developing evidence and insight to design, manage and adapt blockchain-based systems for large-scale tokenised behaviour change to meet diverse goals, whilst avoiding unwanted side-effects such as environmental impact resulting from the platform implementation.

5.2.1 Integrating Targeted Behavioural Changes

Tokens represent value -- stock certificates, a promise for a future product or service (such as a seat in a movie theatre), a proof of ownership, an authorization (permission to operate a motor vehicle), etc. The nascent Token Engineering field has made progress in developing the tools and techniques to design tokens that "behave" in prescribed ways, such that tokens are created (minted), used, and disposed of (burned) according to an ecosystem design.

The envisioned functions for tokens, from real estate, supply chains, data curation/management, etc., require that the expected behaviour of the token be correlated to the expected behaviour of the actors using it. The implication is that as tokens, or more specifically the code that implements and governs tokens, are designed with varying attributes, these attributes should be correlated to Behaviour Change Techniques (BCT), such as the 93 different techniques enumerated by Michie, et al., in the Behaviour Change Technique Taxonomy (2013) and Mechanisms of Action (MA) in order to build evidence on how these tokens achieve their objectives. For example, if tokens are to be used for motivating sustainable travel behaviour, what combination of BCTs and accompanying token design best leverages that behaviour in Vienna versus Lusaka? Does it make a difference if the targeted behaviour is to take public transport, walk or use carpooling? These questions cannot be answered without correlating token design to the human behaviour it targets.

### 5.2.2 A Need for Standardization

Answering questions about which of the analysed sustainable transport research projects was "most effective" would be a frustrating task. Without a certain level of standardization to be able to compare different token designs to effect certain behaviours in certain ecosystems no reliable evidence base can be built. The need to first identify these behaviour change targets (defined by BCTs, MAs, etc.) has been described, but these targets must be correlated to standardized token design patterns to provide for comparison across use cases. The emergent Token Taxonomy Framework is a step in the right direction, as are the technical specifications for token implementation, for example ERC-20[11] and ERC-721[12] on the Ethereum network.

This standardization is critical, especially as more tokenized ecosystems are automated. Without standardization, the research projects outlined here cannot scale and will remain smaller, closed ecosystems that cannot interact with other token platforms. For example, if the Kultur-Token team wanted to expand to include other rewards systems or transportation modalities that have their own tokens, without standardization these platforms would require custom coding to be able to transact with each other.

Practically, tokens are manifested through computer code that provides the governance and economic structures of the ecosystems they manage. The Kultur-Token, for example, has a relationship between behaviour and reward that could be modified at a key stroke if it is found to be no longer be effective. But what evidence guides those keystrokes to change the code (essentially modifying the governance protocols for the ecosystem the token manages)? This evidence base can only be developed by linking behaviourism and tokenism, which requires not only a new sandbox to rapidly develop, deploy, test and adapt but the right skill sets and protocols to do it. As such, an operational framework is needed.

---

[11] https://ethereum.org/en/developers/docs/standards/tokens/erc-20/
[12] https://ethereum.org/en/developers/docs/standards/tokens/erc-721/

### 5.2.3 Operational Frameworks

There have been significant achievements by early innovators in the token design and token economics fields that have laid the groundwork for building this framework. The necessary skill sets have been identified and early lifecycle models have been operationalised (Dhaliwal, et. al, 2019; Voshmgir and Zargham, 2020) but there is no framework that combines the skill sets with a token lifecycle model that is able to meet the objectives of effectively designing for and adapting to behaviour changes in the tokenized ecosystem. As tokens become more sophisticated, their ability to leverage behaviour could exponentially increase. Hence the need to understand this ability requires improved integration of behavioural skill sets, evidence bases and methods into token design, engineering and economics.

## 6    Architecture and Design of Blockchain-based Interventions

The interventions analysed herein typically rely upon evidence of journey being made, which is often a smartphone app that is used to track journeys made by the traveller with the purpose of estimating the distance travelled and the travel mode used, as illustrated in Figure 1. These journey tracking apps, which may also call upon cloud-based servers, use a combination of data captured from sensors on the smartphone and AI/ML technologies. By providing increasingly accurate determinations of journey characteristics, these apps and services are able to provide token economies with an accurate estimation of the carbon saving made by using a travel mode other than a car. Pronello and Kumawat (2020) provide a rich summary of the latest research in this field, which is beyond the scope of this paper.

As work in the C4V project showed, token economies need to be robust to attacks from actors who are motivated to steal from the economy by inflating their income through nefarious means. C4V showed a useful model for identifying and mitigating against potential threats of attack to the economy via cheating on claims of distance travelled or travel mode employed. A further threat that such systems need to mitigate for are attacks on user privacy. Location tracking applications has had a long and difficult history with user privacy, and system architects need to be mindful of the importance of designing their solutions with a privacy first approach, using a methodology such as LINDDUN (Wuyts and Joosen, 2015) to motivate their analysis.

Technically, blockchain platforms are well suited to playing an important role in delivering efficient and secure token economies, and in providing rich solutions which use smart contracts to encode and enforce commitments that users make to each other, or to the environment. Indeed, much research still needs to be done on social and governance design in order to deliver effective and just behaviour change interventions which make best use of the affordances of blockchain technologies. Defining and standardising social and technical interfaces between the actors and components is a critical part of taking blockchain-based systems successfully to scale. Working towards this, Figure 2 provides an initial conceptual architecture for the various components that need careful design, both around their functions and their interactions.

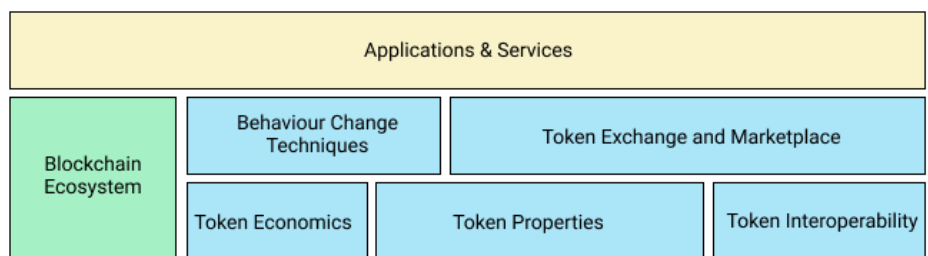

Figure 2 – Components of a blockchain-based behaviour change system

As a case in point, in a closed system such as a purpose-led token economy, the trust that a blockchain infrastructure provides does not complete the story – a blockchain can facilitate secure and immutable transfer of tokens from one party to another, but does not ensure that the marketplace for token conversion will still exist, or that tokens will be given a fair value in that marketplace. Designing the token infrastructure such that the tokens cannot be transferred to other parties, as seen in the C4V project effectively closes the system, and transfers the power of the economy to the marketplace provider. Intriguingly, blockchain platforms can support more open financial models, as exemplified by the decentralised finance (DeFi) movement (Schär, 2020). DeFi provides mechanisms to introduce liquidity into blockchain-based token economies, through platforms such as Uniswap[13] and Balancer[14], and could be leveraged to enable third parties to buy or sell tokens earned in an active travel token economy. A cyclist in Vienna, for example, might prefer to sell their KT tokens in an open market rather than exchange them for a cultural ticket, whereas a city visitor might find that this is provides a cheaper way to obtain a ticket. By providing liquidity to different token economies, DeFi mechanisms can introduce opportunities for arbitrage, and enable the economy to find fair value for earning and exchanging tokens.

The environmental impact of any implemented solution has to be considered at the early stages of any project. Just as the KT and C4V projects consciously considered the impact of their use of blockchain technologies, and so all future projects need to consider the impact of the solution that they are proposing. In particular, the choice of blockchain technology platform needs to be considered carefully, such that the environmental impact of the solution does not weigh heavily on the carbon savings of the desired outcome. Many different blockchain platform implementations are available, and each has a focus which brings strengths and weaknesses in other areas. Energy efficiency may come at the cost of decentralisation, as consensus mechanisms that have lower energy use may involve fewer nodes or use alternative methods that place more control in the hands of a few. The costs associated with transactions on the blockchain platform is also a consideration in any deployment and is determined by the governance model of the blockchain – in some cases costs are fixed, and in others they can be variable, and again there are reasons behind these differences. Research in this area can provide recommendations as to the spread of factors that should be considered in choosing a blockchain platform, such that needs of all stakeholders can be considered and evaluated when project teams make an architectural decision.

## 7   Conclusions

This paper has presented a survey and analysis of projects from the published literature which make use of blockchain technologies to deliver behaviour change interventions with the intent of motivating travellers to make sustainable travel choices. The projects reviewed cover a timespan from 2017 to 2021 and reflect the sociotechnical evolution of blockchain research in that time. It is inspiring to see that the field has developed to the point that city-scale pilot projects such as Kultur-Token can be undertaken.

---

[13] https://uniswap.org
[14] https://balancer.finance

The ongoing challenge of reducing car use in urban environments continues, and the need to change daily travel patterns remains a pressing issue across the globe, where it impacts multiple SDGs. Where individual travellers have a choice in how they make their journeys, behaviour change interventions based upon the development of increasingly complex token economies can have a role to play in influencing fulfilment of that choice. Providing and promoting such initiatives can help city and municipal authorities to maximise the contributions that their citizens make to minimising traffic congestion and air pollution.

As shown, blockchain technology can provide a technical infrastructure capable of facilitating different designs and implementations of tokenised economies to support conversion of journeys to sustainable travel behaviour. However, blockchain technology is just a part of the technical solution, which is reliant on evidence from external sources such as smartphone apps and supporting services that are capable of making informed and accurate judgements on distances users have travelled, whilst balancing motives of data gathering and protection of user privacy. New possibilities brought about by blockchain technology, from autonomous, and individually tailored, programmatic actions of smart contracts and the potential of DeFi to bring liquidity to token economies, offer a tantalising future for delivering engaging programmes to aid travellers in make sustainable choices.

Effective design of behaviour change interventions and their mapping to well-researched token frameworks is critical – determining the actions required of travellers, BCTs employed and governance models all need to be considered by experts with a wide-ranging skill set, guided by best practice and an evidence-base. Blockchain technology and its accompanying ecosystem brings powerful new tools and offers new capabilities for these experts to consider as they design the systems.